\newcommand*{\ATLASLATEXPATH}{latex/}
\pdfoutput=1
\documentclass[USenglish,texlive=2016,cernpreprint,LANGEDIT=false]{\ATLASLATEXPATH atlasdoc}
\usepackage[]{\ATLASLATEXPATH atlaspackage}
\usepackage{\ATLASLATEXPATH atlasbiblatex}
\usepackage{\ATLASLATEXPATH atlasphysics}
\usepackage{\ATLASLATEXPATH atlasheavyion}

\graphicspath{{logos/}{figures/}}
\usepackage{multirow}
\usepackage{dcolumn}

\usepackage{dcolumn} 
\newcolumntype{a}[0]{D{?}{\mbox{--}}{-1}} 
\newcolumntype{b}[1]{D{?}{\pm}{#1} }
\newcolumntype{e}[0]{D{?}{}{-1} }
\newcolumntype{f}[0]{D{?}{>}{-2.1} }

\AtlasTitle{
Observation of centrality-dependent acoplanarity for muon pairs produced via two-photon scattering in Pb+Pb collisions at $\sqrt{s_{\mathrm{NN}}}=5.02~\text{Te\kern -0.1em V}$ with the ATLAS detector
}

\AtlasAbstract{
This Letter presents a measurement of $\gamma \gamma \rightarrow \mu^{+} \mu^{-}$ production in Pb+Pb collisions recorded by the ATLAS detector at the Large Hadron Collider at $\sqrt{s_{\mathrm{NN}}}=5.02~\text{Te\kern -0.1em V}$ with an integrated luminosity of 0.49 nb$^{-1}$.  The azimuthal angle and transverse momentum correlations between the muons are measured as a function of collision centrality.  The muon pairs are produced from $\gamma \gamma$ through the interaction of the large electromagnetic fields of the nuclei. The contribution from background sources of muon pairs is removed using a template fit method.  In peripheral collisions, the muons exhibit a strong back-to-back correlation consistent with previous measurements of muon pair production in ultra-peripheral collisions. The angular correlations are observed to broaden significantly in central collisions. The modifications are qualitatively consistent with rescattering of the muons while passing through the hot matter produced in the collision.
}

\author{The ATLAS Collaboration}

\AtlasRefCode{HION-2018-11}
\PreprintIdNumber{CERN-EP-2018-100}

 \AtlasJournal{Phys.\ Rev.\ Lett.}

\newcommand{\gamgam}{\mbox{$\gamma\gamma$}}

\newcommand{\dZeroPair}{\mbox{$d_{0\,\mathrm{pair}}$}}
\newcommand{\starlight}{\mbox{\textsc{STARlight}}}

\newcommand{\periph}{\mbox{$>$~80\%}}
\newcommand{\gamgammumu}{\ensuremath{\gamma \gamma \rightarrow
    \mumu}\xspace}

\newcommand{\Amumu}{\mbox{$A$}}
\newcommand{\Aco}{\mbox{$\alpha$}}

\newcommand{\kp}{\mbox{$\vec{k}_{\mathrm{T}}$}}

\newcommand{\MSkp}{\mbox{$\left\langle \vec{k}^2_{\scriptscriptstyle \mathrm{T}} \right\rangle$}}

\newcommand{\RMSkp}{\mbox{$k_{\scriptscriptstyle
      \mathrm{T}}^{\scriptscriptstyle \mathrm{RMS}}$}}

\newcommand{\ptbar}{\mbox{$p_{\mathrm{T\,avg}}$}}
\newcommand{\ptbarSq}{\mbox{$p^{2}_{\mathrm{T\,avg}}$}}
\newcommand{\RMSpT}{\mbox{$p_{\scriptscriptstyle \mathrm{T\,
        avg}}^{\scriptscriptstyle \mathrm{RMS}}$}}

\newcommand{\RMSkpValueGaus}{\mbox{$\RMSkp =  66 \pm 10~\MeV$}}
\newcommand{\RMSkpValueConv}{\mbox{$\RMSkp =   70 \pm10~\MeV$}}

\newcommand{\AtlasCoordinateDef}{\footnote{ATLAS uses a
  right-handed coordinate system with its origin at the nominal
  interaction point (IP) in the center of the detector and the
  $z$-axis along the beam pipe. The $x$-axis points from the IP to the
  center of the LHC ring, and the $y$-axis points upward. Cylindrical
  coordinates $(r,\phi)$ are used in the transverse plane, $\phi$
  being the azimuthal angle around the beam pipe. The pseudorapidity
  is defined in terms of the polar angle $\theta$ as
  $\eta=-\ln\tan(\theta/2)$.}}
\newcommand{\templateSignal}{\mbox{$\mathcal{S}$}}
\newcommand{\templateBackground}{\mbox{$\mathcal{B}$}}

\newcommand{\ANpart}{\mbox{$\langle \Npart \rangle $}}
\hypersetup{pdftitle={ATLAS document},pdfauthor={The ATLAS Collaboration}}

\begin{document}

\maketitle

Ultra-relativistic heavy-ion collisions form  hot
strongly interacting matter known as the quark--gluon plasma (QGP)
~\cite{Gyulassy:2004zy,Shuryak:2014zxa,Pasechnik:2016wkt,Busza:2018rrf,}. The
characterization of the properties of the QGP provides unique insight into the dynamics of strongly coupled
many-body systems and is a primary goal of the Relativistic Heavy Ion
Collider (RHIC) and the Large Hadron Collider (LHC) heavy-ion programs.
A common method for studying complex physical systems involves the use
of penetrating probes whose interactions with the system are
well-understood or calibrated. Examples of penetrating probes in heavy-ion collisions are
high-energy quarks and gluons produced in initial hard-scattering
processes~\cite{Qin:2015srf,Mehtar-Tani:2013pia,}. Indeed, many
measurements have shown striking modifications to dijet~\cite{HION-2010-02,HION-2012-11,CMS-HIN-11-013} or
gamma--jet balance~\cite{CMS-HIN-11-010}, the properties of jet-fragment 
distributions~\cite{CMS-HIN-11-004,HION-2012-08,CMS-HIN-12-002,CMS-HIN-12-013,HION-2015-05,CMS-HIN-14-016}, 
and the production rates of high transverse-momentum (\pt) 
hadrons~\cite{CMS-HIN-10-005,HION-2011-03,CMS-HIN-15-015,} or 
jets~\cite{HION-2011-02,HION-2013-06,Adam:2015ewa,CMS-HIN-13-005} as a result of the
interactions of the parent quarks and gluons with the QGP
medium. However, the physics of this phenomenon, known as jet
quenching, is complex, in large part due to the multi-particle nature
of the parton showers that produce the observed jets. 

An alternative, simpler, penetrating probe is provided by $\gamma\gamma
\rightarrow \ell^{+}\ell^{-}$ processes that occur at non-negligible
rates in ultra-relativistic heavy-ion collisions due to the intense
electromagnetic fields generated by ions~\cite{Bertulani:1987tz,Krauss:1997vr,Baur:2001jj,}. The associated photons have
small transverse momenta --- typically less than 10~\MeV\ --- and large
longitudinal momenta and energies~\cite{Cahn:1990jk,Klein:2016yzr}. For example, in $\sqn = 5.02$~\TeV\ \PbPb\
collisions at the LHC, the photon energy spectra and the resulting
lepton \pT\ distributions extend to about $50$~\GeV. Due to the low
transverse momenta of the photons, the leptons are produced nearly back-to-back
in azimuth and with nearly identical transverse momenta.
Photon-induced scattering processes in heavy-ion collisions are
typically studied in so-called ultra-peripheral collisions (UPCs)~\cite{Baltz:2007kq,Bertulani:2005ru} 
for which the impact parameter
between the colliding nuclei is larger than twice the nuclear radius,
such that there is no hadronic interaction between the
nuclei.  UPC events are used to study exclusive
vector-meson production in photon--nucleus collisions~\cite{Abelev:2007nb,Agakishiev:2011me,Adam:2015gsa,Afanasiev:2009hy,Abelev:2012ba,Abbas:2013oua,Adam:2015sia},
lepton-pair production in photon--photon collisions~\cite{Abbas:2013oua}, and
recently, light-by-light scattering~\cite{Aaboud:2017bwk}.

Although photon-induced reactions are typically measured in UPCs, they
have also been observed in hadronic collisions of heavy ions \cite{Adam:2015gba,Adam:2018tdm}. In such events, the
photon fluxes are largest just outside the nuclear overlap region, and
it is expected that charged leptons produced in this region interact with
the electric charges in the QGP that is formed. While the
effects of electromagnetic interactions are much weaker than the
strong interactions responsible for jet quenching, the initial angles and momenta
of the produced leptons are sufficiently well correlated to make 
even small modifications observable. One potential source of 
modification is the final-state interaction of the produced leptons
with the electric charges in the QGP. In this scenario, small momentum
transfers to the leptons due to electromagnetic 
interactions may result in the broadening of the momentum and angular
correlations of the lepton pair, in analogy with the original picture of
jet energy loss proposed by Bjorken~\cite{Bjorken:1982tu}. Such broadening should be largest in central
collisions, where the degree of overlap between the colliding nuclei
is greatest and the transverse size and lifetime of the plasma are largest.
Unlike jet observables~\cite{HION-2010-02,CMS-HIN-10-004,CMS-HIN-11-013,HION-2012-11},
measurements using lepton pairs in this fashion have not been
explored previously. Jets are multi-particle systems
consisting of a shower of quarks and gluons. Measurements of the
modification of these showers provide detailed information about the
microscopic structure of the QGP over a range of length scales but at
the expense of introducing significant complexity to the problem. The
interaction of lepton pairs with the medium is much simpler, and thus
measurements using such pairs are a critical baseline for
understanding jet quenching.

This Letter reports a measurement by ATLAS of the angular and momentum
correlations of muon pairs produced via photon--photon scattering in
5.02~\TeV\ \PbPb\ collisions  using data with an
integrated luminosity of 0.49~\inb\ recorded during the 2015 \PbPb\
operation of the LHC. The
$\gamma\gamma \rightarrow \mu^+\mu^-$ pairs are distinguishable 
from muon pairs arising from other production mechanisms through their
angular and momentum correlations, which are quantified using the pair acoplanarity, \Aco,
and asymmetry, \Amumu, defined as:
\begin{equation}
\Aco\equiv 1-\frac{|\phi^+ - \phi^-|}{\pi}\,,\quad \Amumu \equiv \left |\frac{ \pt^{+}
  - \pt^{-}}{\pt^{+} + \pt^{-}}\right|\,, \nonumber
\end{equation}
where $\phi^{\pm}$ represent the azimuthal angles and
$p_{\mathrm{T}}^{\pm}$ the magnitudes of the transverse momenta of the two muons. The distributions of these
quantities from $\gamma\gamma \rightarrow \mu^+\mu^-$ pairs are
extremely peaked near zero due to the small transverse momentum of
the \gamgam\ system.
Background at small \Aco\ and \Amumu, 
resulting from semileptonic
decays of heavy-flavor (HF) hadrons, is subtracted using a template fit method exploiting the 
fact that these hadrons often decay after traveling a significant
distance from the 
interaction point. Other background
contributions such as Drell--Yan and $\Upsilon$ production and
dissociative processes~\cite{Vermaseren:1982cza} are observed to be
negligible over the narrow range of \Aco\ and \Amumu\ considered
here. The \Aco\ and \Amumu\ distributions are presented 
for different intervals of \PbPb\ collision centrality. A broadening
observed in the \Aco\ distributions is characterized using a
fitting procedure that provides a transverse momentum scale, \RMSkp.

The data are recorded with the ATLAS detector~\cite{PERF-2007-01} 
using its calorimeter, inner detector, muon
spectrometer, trigger, and data acquisition
systems.\AtlasCoordinateDef\ The calorimeter system consists of a
liquid-argon (LAr) electromagnetic calorimeter covering $|\eta|<3.2$, a
steel/scintillator sampling hadronic calorimeter covering $|\eta| <
1.7$, a LAr hadronic calorimeter covering $1.5 < |\eta| < 3.2$, and a
forward calorimeter (FCal) covering $3.2 < |\eta| <
4.9$. Charged-particle tracks are measured over the range $|\eta|<2.5$
using the inner detector, which is composed of
silicon pixel detectors in the innermost layers, followed by silicon
microstrip detectors and a straw-tube transition-radiation tracker
($|\eta| < 2.0$), all immersed in a 2~T axial magnetic field. The muon
spectrometer system comprises separate trigger and high-precision
tracking chambers, covering $|\eta| < 2.4$ and $|\eta| < 2.7$,
respectively, measuring the deflection of muons in a magnetic
field provided by superconducting air-core toroid magnets.

Events used in this measurement are
selected by a trigger requiring at least two muons~\cite{TRIG-2016-01}, each having $\pt >
4$~\GeV. Events are further required to have a
reconstructed primary vertex, built from at least two tracks with $\pt
> 0.4$~\GeV. The collision centrality is determined by
analyzing the total transverse energy measured in the FCal 
in minimum-bias \PbPb\ collisions and dividing the distribution into
centrality intervals corresponding to successive quantiles of the
total~\cite{Miller:2007ri}. 
The intervals used in
this measurement are 0--10\%, 10--20\%, 20--40\%, 40--80\%, and \periph,
which are ordered from the most central (highest transverse energy) to
most peripheral. The \periph\ interval includes the 80--100\%
centrality interval as well as UPC events, which
contain most of the muon pairs measured in that interval.

The detector response to signal muon pairs is evaluated using Monte Carlo
(MC) samples of 
 $\ce{Pb}+\ce{Pb}\rightarrow \ce{Pb^{(*)}}  \gamma
 \gamma \ce{Pb^{(*)}} \rightarrow  \ce{Pb^{(*)}} \mu^{+}\mu^{-} \ce{Pb^{(*)}}$ events,
produced with the \starlight\ event
generator~\cite{Klein:2016yzr,Baltz:2009jk}, which utilizes the
equivalent photon approximation and neglects the initial transverse
momentum of the incoming photons. A separate MC sample of
background muon pairs resulting from heavy-flavor decays was produced using
\PYTHIA 8.185~\cite{Sjostrand:2007gs} with the A14 set of tuned parameters~\cite{ATL-PHYS-PUB-2014-021}
and NNPDF2.3 LO parton distribution functions~\cite{Ball:2012cx}. Both samples
were passed through a \GEANT4\ \cite{Agostinelli:2002hh} simulation of the detector and overlaid on 
minimum-bias \PbPb\ data. The resulting events were reconstructed
in the same manner as the data.  

The analysis is performed by considering all oppositely charged muon
pairs in the events meeting the trigger and event selection
requirements. The muons are identified by matching tracks in
the muon spectrometer to tracks in the inner detector. Each muon is required to have $\pt > 4$~\GeV\ and $|\eta|
< 2.4$~\cite{PERF-2014-05,PERF-2015-10}. An invariant mass requirement of $4 < m_{\mu^+\mu^-} < 45$~\GeV\ is
applied to suppress the contribution from hadron (primarily \Jpsi)
decays and $\Zzero$ boson decays to muon pairs. In order to
account for inefficiency introduced by the trigger and reconstruction,
each muon is weighted by $w=(\varepsilon_{\mathrm{trig}}\,
\varepsilon_{\mathrm{reco}})^{-1}$ when constructing the
distributions. Both efficiencies are functions of the muon $\pt$ and
$\eta$ and are obtained from studies of $\Jpsi\rightarrow\mumu$
decays~\cite{HION-2014-05,HION-2016-07}. The efficiencies rise rapidly as a
function of \pT\ before reaching constant values of approximately 0.8
to 0.95 for $\pT > 5$~\GeV, depending on the $\eta$ value. Systematic
uncertainties due to the efficiency corrections are evaluated by
varying each efficiency by its uncertainty. These variations have little
impact on the measurement since they largely cancel out in final
observables, which are normalized by the total yield.

The \Aco\ and \Amumu\ distributions include significant background
from HF decays. The background \Aco\ and \Amumu\ distributions are
each obtained from data by making selections on the other variable to
suppress the \gamgam\ contribution. Specifically, the background
\Aco\ distribution is constructed by requiring $\Amumu >
0.06$, and the background \Amumu\ distribution is obtained by requiring
$\Aco > 0.015$.  These selections were found not
to significantly alter the distributions in the HF MC sample. 
In order to minimize the influence of statistical fluctuations,
both the background \Aco\ and \Amumu\ distributions were assumed to be smooth functions,
determined by fitting them with second-order polynomials.
Systematic uncertainties in the shapes of these
distributions are evaluated by propagating statistical uncertainties
obtained from the fits including covariance between the parameters.
The systematic uncertainty of the background shape is evaluated by performing the
fits with linear and constant functions.  

\begin{figure*}[t]
\centering
\includegraphics[width=0.95\textwidth]{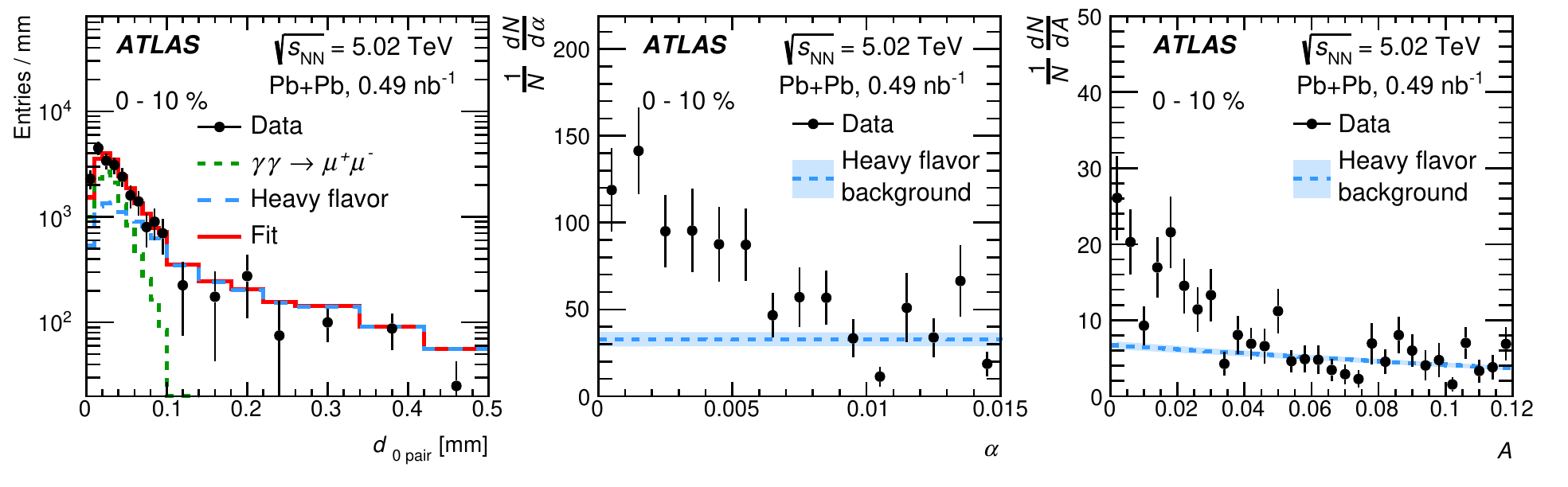}
\caption{Template fits (left) to the \dZeroPair\ distributions are
  shown for the
  0--10\% interval. The  \Aco\ (center) and
\Amumu\ (right) distributions are shown before background subtraction (points). These distributions are normalized to unity
over their measured range. In the central and right plots, the background
contributions with normalization fixed by the template fitting are
indicated by the dashed line with the uncertainties represented by the
shaded band.}
\label{fig:template_fitting}
\end{figure*}

The normalization of the background \Aco\ and \Amumu\ distributions is
determined using a template-fitting procedure.
The quadrature sum $\dZeroPair\equiv d_{0}^{+} \oplus d_{0}^{-}$ is
constructed for muon each pair, where $d_{0}^{\pm}$ are the
transverse impact parameters of the track trajectories of the
individual muons relative to the
collision vertex. The template fitting is
performed over the signal-enriched kinematic
range $\Aco <0.015$ and $\Amumu < 0.06$.
The \dZeroPair\ distributions are fit using the function $\mathcal{F}(\dZeroPair)\equiv f\templateSignal(\dZeroPair) +
(1-f)\templateBackground(\dZeroPair)$, where \templateSignal\ and
\templateBackground\ are the \gamgam\ signal and HF background distributions,
respectively, to obtain the signal fraction, $f$. The
\templateSignal\ distributions are determined primarily by multiple
scattering and detector resolution,
and are obtained from the \starlight\ MC sample. The \templateBackground\
distributions have long tails as one or both
of the HF hadrons may travel a significant distance before decaying. 
These \templateBackground\ distributions are obtained from data by requiring that $\Amumu > 0.15$ and
$\Aco > 0.02$. Since the signal process populates only small values of
\Amumu\ and \Aco, the \templateBackground\ distributions obtained in this way are
dominated by the HF contribution in the data. In the 40--80\% 
and \periph\ centrality intervals, the distributions from the
HF MC sample were used, as the data did not contain enough events 
after applying these selections to construct a template. An example of 
the template fitting for the 0--10\%
centrality interval is shown in the left panel of Figure~\ref{fig:template_fitting}.
Uncertainties in the signal fractions resulting from the
\templateSignal\ shape are obtained by modifying the fit function, $\mathcal{F}_{\mathrm{sys}} (\dZeroPair)
\equiv f\templateSignal(c \dZeroPair) + (1-f)\templateBackground(\dZeroPair)$, where $c$
is an additional free parameter in the fitting that enables scaling
of the \templateSignal\ distributions along the \dZeroPair\ axis; this variation
accounts for possible inaccuracies in the $d_0$ resolution in the
\starlight\ MC sample. Uncertainties due to the 
\templateBackground\ template are evaluated by varying the
requirements on \Aco\ and \Amumu\ in the definition of the background
region. The signal fraction in the 0--10\% interval is $f=0.51\pm
0.03$, and generally increases in more peripheral collisions, becoming 
consistent with no background contribution in the most peripheral
interval, \periph.

The \Aco\ and \Amumu\ distributions are obtained from the data by
restricting the range of the other variable: $\Amumu < 0.06$ and $\Aco 
< 0.015$, respectively. Both the data obtained in this fashion and the background distributions
are shown in the center and right panels of Figure~\ref{fig:template_fitting}
respectively, for the 0--10\%
centrality interval. 

\begin{figure*}[t]
\centering
\includegraphics[width=0.95\textwidth]{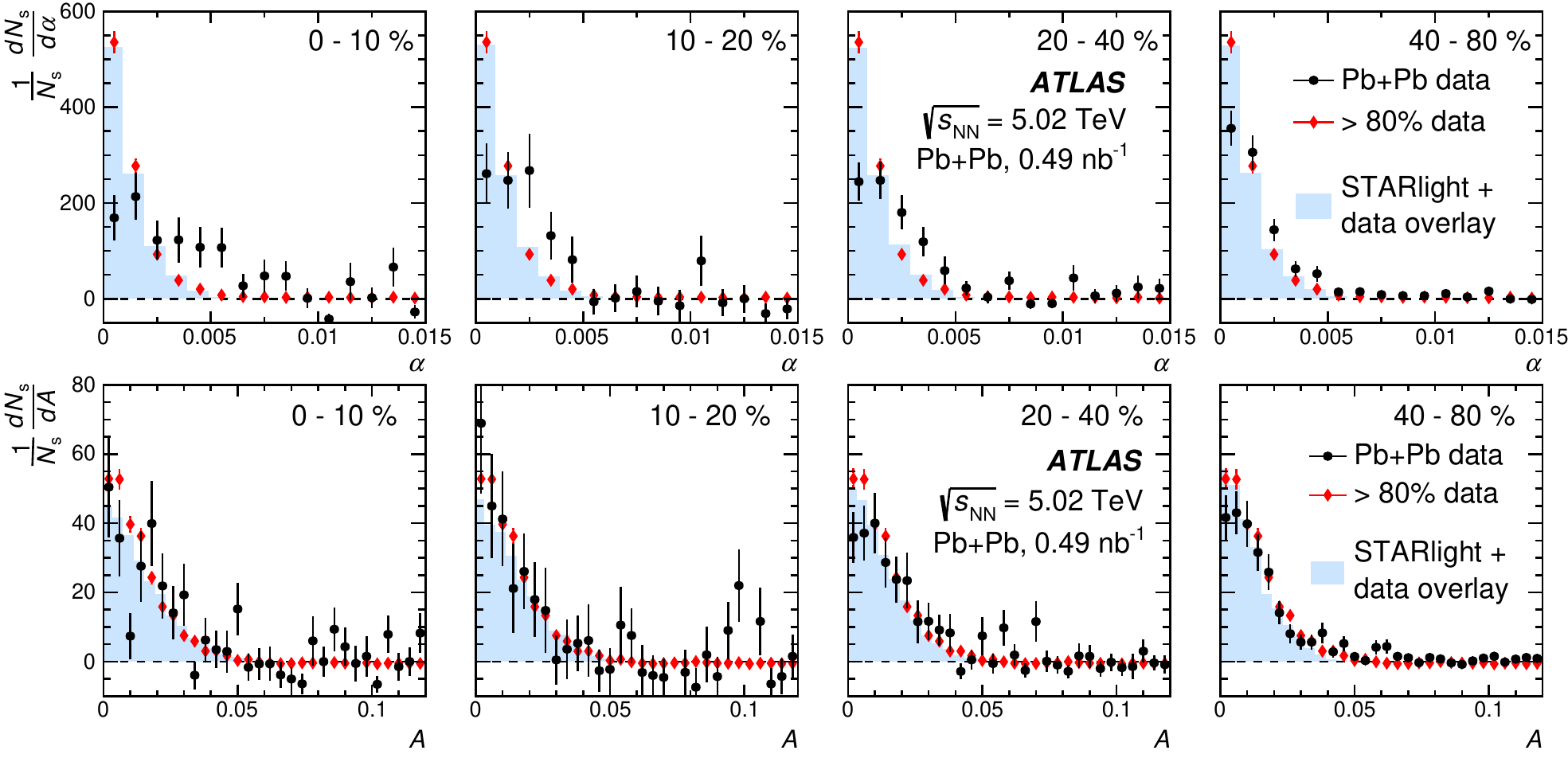}
\caption{
The background-subtracted distributions are shown for \Aco\
(upper row) and \Amumu\ (lower row).  Each distribution is normalized to
unity over its measured range. Moving from left to right, the data
(circles) are shown for increasingly peripheral collisions (lower degree
of overlap, higher percentile). The distributions obtained from the MC
simulation (\gamgammumu\ generated by \starlight\ and overlaid on data)
are shown for the corresponding centrality interval as a filled
histogram. The distribution measured in the most peripheral
collisions, the \periph\ interval (diamonds)
is repeated in each panel to facilitate a direct comparison. The error
bars include the statistical and systematic
uncertainties. Uncertainties related to the background normalization
are not shown.}

\label{fig:results}
\end{figure*}
The background-subtracted distributions
$(1/N_{\mathrm{s}}) dN_{\mathrm{s}}/d\alpha$ and $(1/N_{\mathrm{s}}) dN_{\mathrm{s}}/dA$ 
measured in different
centralities in the data are shown in Figure~\ref{fig:results} in the
top and bottom rows, respectively. Each distribution is normalized to
unity over its measured range. The \periph\ distribution is plotted 
in each panel for comparison. The systematic uncertainties affecting the
background normalization and shape are not shown in this figure. These uncertainties
are generally negligible compared with the statistical uncertainties
indicated by the error bars, and they exhibit strong correlations as a
function of \Aco\ and \Amumu. After background subtraction, both data distributions are
consistent with zero at the largest values of \Aco\ and \Amumu\
considered in the measurement. This feature indicates that other sources of
background, such as Drell--Yan and $\Upsilon$ production and
dissociative processes, which are
essentially constant over the measurement range, are not a significant
contribution. A clear, centrality-dependent broadening
is seen in the acoplanarity distributions when compared to the \periph\
interval. No such effect is seen for the asymmetry distributions.  The corresponding
distributions from the \gamgammumu\ MC samples are also shown. The MC \Aco\ distributions show almost no centrality
dependence, indicating that the broadening evident in the data is
notably larger than that expected from detector effects. Although the
\Amumu\ distributions from the MC sample broaden slightly in more
central collisions, they are intrinsically much broader than the corresponding \Aco\
distributions. 

\begin{figure*} \centering
\includegraphics[width=0.95\textwidth]{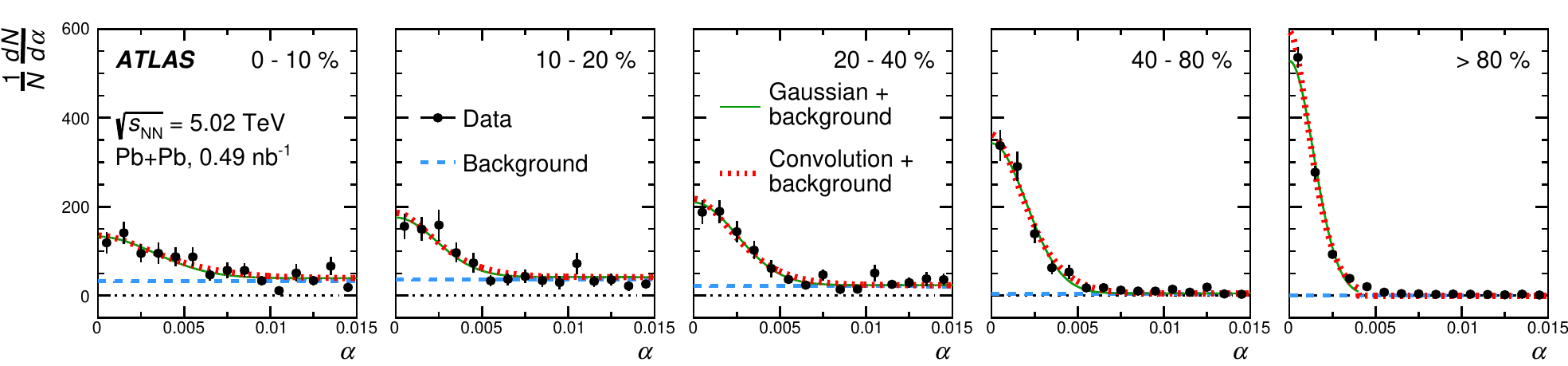}
 \caption{Results of fits to the muon pair \Aco\ distributions using 
the sum of Gaussian and background functions. A standard Gaussian function is
shown as a solid curve while the dotted curve shows a Gaussian function in
\Aco\ convolved with the measured \ptbar\ distribution. The background
distributions are indicated by the dashed lines. }
\label{fig:kt_fits}
\end{figure*}

In order to quantify the broadening observed in the \Aco\ distributions, the unsubtracted
distributions are fit to a Gaussian function plus the
normalized background distribution. The fit functions are shown with
the solid curves in Figure~\ref{fig:kt_fits} and the values of the
width, $\sigma$,
are listed in Table~\ref{tbl:RMSAcoAsymm}. The $\sigma$ values 
increase by more than a factor of two between the most peripheral interval and the most central interval.  
Similar fits are performed for the \Amumu\ distributions and 
the resulting $\sigma$ values are listed 
in Table~\ref{tbl:RMSAcoAsymm}. Unlike the \Aco\ distributions, no 
significant broadening of the \Amumu\ distributions can be inferred.
\renewcommand{\arraystretch}{1.4}
\begin{table}
\caption{Values of the parameters obtained by applying the Gaussian and
  convolution fits to the \Aco\ distributions shown in Fig.~\ref{fig:kt_fits} for different intervals of
  centrality. Also shown are the average number of participants, \ANpart; the RMS
  \ptbar, \RMSpT, used to relate the $\sigma$ parameter to \RMSkp\ in
  the Gaussian fitting procedure; and the $\sigma$ parameter obtained
  from applying the Gaussian fitting to the \Amumu\ distributions.}
\centerline{
\begin{tabular}{|r|b{-1}|b{1.1}|e|b{1.1}|b{1.1}|b{1.1}|} \hline
\multicolumn{1}{|r|}{\multirow{2}{*}{Centrality}} & \multicolumn{1}{c|}{\multirow{2}{*}{\ANpart}} &  \multicolumn{1}{c|}{\multirow{2}{*}{\RMSpT\ [\GeV]}}&\multicolumn{3}{c|}{Gaussian fit} & \multicolumn{1}{c|}{Convolution fit} \\ \cline{4-7}
&  &   & \multicolumn{1}{c|}{$\sigma_{A} (\times 10^3) $} & \multicolumn{1}{c|}{$\sigma_{\alpha} (\times 10^3)$} & \multicolumn{1}{c|}{\RMSkp\ [\MeV]}& \multicolumn{1}{c|}{\RMSkp\ [\MeV]}\\ \hline
  0--10\% &       359?2 &            7.0?0.1 &           17.9? \genfrac{}{}{0pt}{}{+1.0}{-0.9} &            3.3? 0.4 &             66? 10 &             70? 10 \\ \hline
    10--20\% &       264?3 &            7.7?0.4 &           13.6? \genfrac{}{}{0pt}{}{+1.2}{-1.0} &            2.3? 0.3 &             40? 7 &             42? 7 \\ \hline
    20--40\% &       160?3 &            7.4?0.3 &           17.2? \genfrac{}{}{0pt}{}{+0.4}{-0.4} &            2.5? 0.2 &             48? 6 &             44? 5 \\ \hline
    40--80\% &        47?2 &            6.8?0.3 &           16.1? \genfrac{}{}{0pt}{}{+0.1}{-0.1} &            2.0? 0.1 &             35? 4 &             32? 2 \\ \hline
  $>$ 80\% &\multicolumn{1}{c|}{--} &            7.0?0.3 &           15.5?\genfrac{}{}{0pt}{}{+0.1}{-0.1} &           1.40?0.03 &\multicolumn{1}{c|}{-} &\multicolumn{1}{c|}{-}\\ \hline
\end{tabular}
}
\label{tbl:RMSAcoAsymm}
\end{table}

Assuming that the
broadening of the \Aco\ distributions results from a physical process
that transfers a {\em small} amount of transverse 
momentum, $|\kp| \ll p^{\pm}_{\mathrm{T}}$, to each muon then
the variance of the \Aco\ distribution can be approximated as
\begin{equation}
\langle \Aco^2 \rangle
= \langle \Aco^2 \rangle_{0}
+ \frac{1}{\pi^2}\frac{\MSkp}{\left\langle \ptbarSq\right\rangle}\,,
\label{eq:acokt}
\end{equation}
where \ptbar\ is the average of $\pt^{+}$ and $\pt^{-}$ and $\langle \Aco ^2 \rangle_{0}$ is the
intrinsic mean square acoplanarity resulting from both the production process
itself and the angular resolution in the muon measurement.

Taking $\langle \Aco^2 \rangle_0$ to be the $\sigma^2$ of the Gaussian
fit in the \periph\ interval, an estimate of the root-mean-square
(RMS) of
$|\kp|$, \RMSkp, is evaluated in each centrality interval
using the measured value of the RMS value of \ptbar,  and substituting
 $\sigma^2$ of the Gaussian fit in that centrality interval for $\langle
 \Aco^2 \rangle$. For the 0--10\% centrality interval this procedure gives \RMSkpValueGaus.

The variance of the \Amumu\ distribution obeys a relation similar to
Eq.~(\ref{eq:acokt}) but with $1/\pi^2$ substituted by  $1/4$. If the
values obtained above for \RMSkp\ are used in that equation an increase
of only about 0.001 in the RMS of \Amumu\ is expected between \periph\ and
0--10\% collisions. The insensitivity of the asymmetry to the
broadening observed in the acoplanarity distributions can be
understood as resulting from the roughly five times larger intrinsic
width of the \Amumu\ distribution. This larger width is consistent
with, and can be attributed to, the momentum resolution of the ATLAS
inner detector~\cite{STDM-2015-02}.  

\begin{figure} \centering
\includegraphics[width=0.49\textwidth]{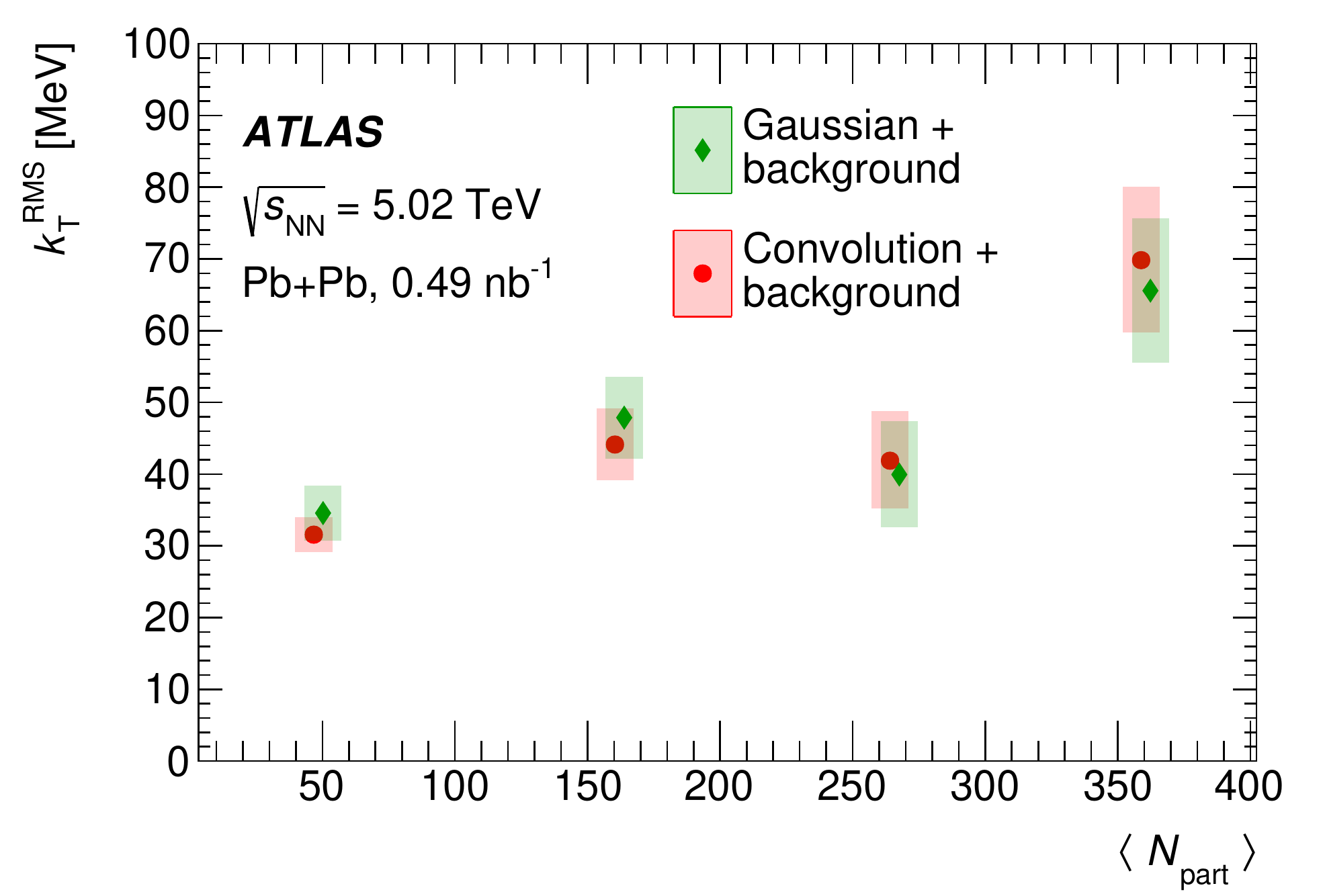}
 \caption{The \RMSkp\ values obtained from the fits shown in
   Figure~\ref{fig:kt_fits} as a function of $\langle
   N_{\mathrm{part}} \rangle$. The shaded bands indicate the
 total uncertainty accounting for both the systematic and statistical
 uncertainties in the \Aco\ distributions and background. The data points have been horizontally 
 offset for visualization purposes, and the horizontal sizes of the
 error bands are arbitrary. }
\label{fig:rmskt}
\end{figure}

This fitting procedure provides a direct relationship between the widths of
the \Aco\ distributions and the \RMSkp\ but does not fully account for
the shape of the \ptbar\ distributions. This limitation is addressed by an
alternative procedure, in which the unsubtracted \Aco\ distributions are
fit as above but replacing the
Gaussian function with a function produced by
convolving the measured \ptbar\ distribution in each centrality
interval with a Gaussian function in \Aco\ of width $\sqrt{ (\RMSkp)^{\scriptsize{\,2}}+k_{\scriptscriptstyle
      \mathrm{T}\,0}^{\scriptscriptstyle 2}}/\pi \ptbar$. The
parameter $k_{\scriptscriptstyle \mathrm{T}\,0}$ is obtained from the
fit to the data in the \periph\ centrality interval. 
The results of these fits are also shown in Figure~\ref{fig:kt_fits}, and 
the obtained \RMSkp\ values are shown 
in Figure~\ref{fig:rmskt} as a function of \ANpart, the average number of
participant nucleons in each centrality interval obtained from a Glauber
model analysis~\cite{Miller:2007ri}. Also shown in
Figure~\ref{fig:rmskt} are estimates for
\RMSkp\ obtained by applying
Eq.~(\ref{eq:acokt}) to the results of the Gaussian acoplanarity fits. The two methods yield results that
are consistent within their uncertainties. With both methods,
the extracted \RMSkp\ is observed to grow from more peripheral to
more central collisions, or equivalently, from smaller to larger \ANpart.
In the 0--10\% centrality interval \RMSkpValueConv. Variations of the 
\ptbar-convolution fitting are also performed allowing an additional
background contribution consistent with Drell--Yan and dissociative
processes. The extracted \RMSkp\ agree with those
reported here well within the
uncertainties associated with the background subtraction. Although
the discussion here formulates the broadening as momentum transfer
applied to each muon, the analysis does not assume that such final-state effects are the
only possible mechanism, and the physical interpretation of the \RMSkp\
values is not limited to this paradigm. More generally, the \RMSkp\
values provide an estimate of a transverse momentum scale associated
with a physical mechanism absent in the typical description of coherent
\gamgam\ processes in heavy-ion collisions, in which the \gamgam\
system has much less initial transverse momentum.

In conclusion, this Letter reports a measurement of muon pair production
in \PbPb\ collisions in which the pairs are produced
electromagnetically through the process \gamgammumu. Contributions
from heavy-flavor decays are removed and the
resulting \Aco\ and \Amumu\ distributions exhibit a
strong correlation attributable to the small transverse momentum of
the initial \gamgam\ system. The \Aco\ distributions are observed to
broaden in increasingly central collisions. No such broadening is seen 
in the \Amumu\ distributions, where the sensitivity is limited by
momentum resolution. A transverse momentum
scale quantifying the magnitude of the broadening, relative to \periph\ collisions, is extracted from
the \Aco\ distributions. In the 0--10\% centrality interval, that scale,
assumed to be the RMS momentum transfer to each final-state muon in
the transverse plane, is \RMSkpValueConv.

\section*{Acknowledgments}

We thank CERN for the very successful operation of the LHC, as well as the
support staff from our institutions without whom ATLAS could not be
operated efficiently.

We acknowledge the support of ANPCyT, Argentina; YerPhI, Armenia; ARC, Australia; BMWFW and FWF, Austria; ANAS, Azerbaijan; SSTC, Belarus; CNPq and FAPESP, Brazil; NSERC, NRC and CFI, Canada; CERN; CONICYT, Chile; CAS, MOST and NSFC, China; COLCIENCIAS, Colombia; MSMT CR, MPO CR and VSC CR, Czech Republic; DNRF and DNSRC, Denmark; IN2P3-CNRS, CEA-DRF/IRFU, France; SRNSFG, Georgia; BMBF, HGF, and MPG, Germany; GSRT, Greece; RGC, Hong Kong SAR, China; ISF, I-CORE and Benoziyo Center, Israel; INFN, Italy; MEXT and JSPS, Japan; CNRST, Morocco; NWO, Netherlands; RCN, Norway; MNiSW and NCN, Poland; FCT, Portugal; MNE/IFA, Romania; MES of Russia and NRC KI, Russian Federation; JINR; MESTD, Serbia; MSSR, Slovakia; ARRS and MIZ\v{S}, Slovenia; DST/NRF, South Africa; MINECO, Spain; SRC and Wallenberg Foundation, Sweden; SERI, SNSF and Cantons of Bern and Geneva, Switzerland; MOST, Taiwan; TAEK, Turkey; STFC, United Kingdom; DOE and NSF, United States of America. In addition, individual groups and members have received support from BCKDF, the Canada Council, CANARIE, CRC, Compute Canada, FQRNT, and the Ontario Innovation Trust, Canada; EPLANET, ERC, ERDF, FP7, Horizon 2020 and Marie Sk{\l}odowska-Curie Actions, European Union; Investissements d'Avenir Labex and Idex, ANR, R{\'e}gion Auvergne and Fondation Partager le Savoir, France; DFG and AvH Foundation, Germany; Herakleitos, Thales and Aristeia programmes co-financed by EU-ESF and the Greek NSRF; BSF, GIF and Minerva, Israel; BRF, Norway; CERCA Programme Generalitat de Catalunya, Generalitat Valenciana, Spain; the Royal Society and Leverhulme Trust, United Kingdom.

The crucial computing support from all WLCG partners is acknowledged gratefully, in particular from CERN, the ATLAS Tier-1 facilities at TRIUMF (Canada), NDGF (Denmark, Norway, Sweden), CC-IN2P3 (France), KIT/GridKA (Germany), INFN-CNAF (Italy), NL-T1 (Netherlands), PIC (Spain), ASGC (Taiwan), RAL (UK) and BNL (USA), the Tier-2 facilities worldwide and large non-WLCG resource providers. Major contributors of computing resources are listed in Ref.~\cite{ATL-GEN-PUB-2016-002}.

\printbibliography

\clearpage 
 
\begin{flushleft}
{\Large The ATLAS Collaboration}

\bigskip

M.~Aaboud$^\textrm{\scriptsize 34d}$,
G.~Aad$^\textrm{\scriptsize 99}$,
B.~Abbott$^\textrm{\scriptsize 124}$,
O.~Abdinov$^\textrm{\scriptsize 13,*}$,
B.~Abeloos$^\textrm{\scriptsize 128}$,
D.K.~Abhayasinghe$^\textrm{\scriptsize 91}$,
S.H.~Abidi$^\textrm{\scriptsize 164}$,
O.S.~AbouZeid$^\textrm{\scriptsize 39}$,
N.L.~Abraham$^\textrm{\scriptsize 153}$,
H.~Abramowicz$^\textrm{\scriptsize 158}$
% ...
\bigskip
\\

% ...
$^{13}$Institute of Physics, Azerbaijan Academy of Sciences, Baku; Azerbaijan.\\
% ...
$^{34}$$^{(a)}$Facult\'e des Sciences Ain Chock, R\'eseau Universitaire de Physique des Hautes Energies - Universit\'e Hassan II, Casablanca;$^{(b)}$Centre National de l'Energie des Sciences Techniques Nucleaires (CNESTEN), Rabat;$^{(c)}$Facult\'e des Sciences Semlalia, Universit\'e Cadi Ayyad, LPHEA-Marrakech;$^{(d)}$Facult\'e des Sciences, Universit\'e Mohamed Premier and LPTPM, Oujda;$^{(e)}$Facult\'e des sciences, Universit\'e Mohammed V, Rabat; Morocco.\\
% ...
$^{39}$Niels Bohr Institute, University of Copenhagen, Copenhagen; Denmark.\\
% ...
$^{91}$Department of Physics, Royal Holloway University of London, Egham; United Kingdom.\\
% ...
$^{99}$CPPM, Aix-Marseille Universit\'e, CNRS/IN2P3, Marseille; France.\\
% ...
$^{124}$Homer L. Dodge Department of Physics and Astronomy, University of Oklahoma, Norman OK; United States of America.\\
% ...
$^{128}$LAL, Universit\'e Paris-Sud, CNRS/IN2P3, Universit\'e Paris-Saclay, Orsay; France.\\
% ...
$^{153}$Department of Physics and Astronomy, University of Sussex, Brighton; United Kingdom.\\
% ...
$^{158}$Raymond and Beverly Sackler School of Physics and Astronomy, Tel Aviv University, Tel Aviv; Israel.\\
% ...
$^{164}$Department of Physics, University of Toronto, Toronto ON; Canada.\\
% ...
$^{a}$ Also at Borough of Manhattan Community College, City University of New York, New York City; United States of America.\\
$^{b}$ Also at Centre for High Performance Computing, CSIR Campus, Rosebank, Cape Town; South Africa.\\
$^{c}$ Also at CERN, Geneva; Switzerland.\\
$^{d}$ Also at CPPM, Aix-Marseille Universit\'e, CNRS/IN2P3, Marseille; France.\\
$^{e}$ Also at Departament de Fisica de la Universitat Autonoma de Barcelona, Barcelona; Spain.\\
% ...
$^{*}$ Deceased

\end{flushleft}

 
\end{document}